\documentclass[sigconf, nonacm]{acmart}

\usepackage{bm}
\usepackage{multirow}
\usepackage{amsmath}
\usepackage{url}
\usepackage{hyperref}
\usepackage{multirow}
\usepackage{epsfig}
\usepackage{epstopdf}
\usepackage{graphicx}
\graphicspath{{./figures/}} 

\usepackage{balance}

\begin{document}

\title{Large Scale Record Linkage in the Presence of Missing Data}

\author{Thilina Ranbaduge}
\orcid{1234-5678-9012}
\affiliation{%
  \institution{The Australian National University}
  \city{Canberra}
  \country{Australia}
}
\email{thilina.ranbaduge@anu.edu.au}

\author{Peter Christen}
\orcid{0000-0003-3435-2015}
\affiliation{%
  \institution{The Australian National University}
  \city{Canberra}
  \country{Australia}
 }
\email{peter.christen@anu.edu.au}

\author{Rainer Schnell}
\orcid{0000-0001-7843-4974}
\affiliation{%
  \institution{University Duisburg-Essen}
  \city{Duisburg}
  \country{Germany}
}
\email{rainer.schnell@uni-due.de}


\begin{abstract}
Record linkage is aimed at the accurate and efficient
identification of records that represent the same entity within
or across disparate databases. It is a fundamental task in data
integration and increasingly required for accurate decision making
in application domains ranging from health analytics to national
security. Traditional record linkage techniques calculate string
similarities between quasi-identifying (QID) values, such as the
names and addresses of people. Errors, variations, and missing
QID values can however lead to low linkage quality because the
similarities between records cannot be calculated accurately. To
overcome this challenge, we propose a novel technique that can
accurately link records even when QID values contain errors or
variations, or are missing. We first generate attribute signatures
(concatenated QID values) using an Apriori based selection of
suitable QID attributes, and then relational signatures that
encapsulate relationship information between records. Combined,
these signatures can uniquely identify individual records and
facilitate fast and high quality linking of very large databases
through accurate similarity calculations between records. We
evaluate the linkage quality and scalability of our approach using
large real-world databases, showing that it can achieve high
linkage quality even when the databases being linked contain
substantial amounts of missing values and errors.
\end{abstract}


\maketitle



\section{Introduction}

Organisations such as financial institutions, statistical agencies,
and government departments, increasingly require records about
entities from multiple databases to be integrated to allow efficient
and accurate decision making~\cite{Dong2015morgan},
where such databases can contain many millions of records with detailed
information about people, such as customers or
patients~\cite{Chr12a,Get15}.
%
\emph{Record linkage} (RL)~\cite{Fel69} is one major task required
when databases are to be integrated. RL aims to 
identify and match records that refer to the same entities across 
different databases~\cite{Chr12a}. Linked databases allow improvement
of data quality, enrichment of the information known about entities,
and facilitate the discovery of novel patterns and relationships
between entities that cannot be identified from individual
databases~\cite{Christen2020lsd,Dong2015morgan,Wan11}.

Because there is often a lack of unique entity identifiers (such as
social security numbers) across the databases to be linked, RL is 
commonly based on partially identifying attributes, known as 
\emph{quasi-identifiers} (QIDs)~\cite{Chr13b}, such as names, 
addresses, dates of birth, and so on. Data quality aspects such as 
typographical errors, variations, and changes of values over time 
(for example when people move or change their names due to marriage)
are common in many of the QID attributes used for RL~\cite{Chr12a}.
Due to errors and variations in QIDs, exact matching of attribute
values can lead to poor linkage quality~\cite{Chr12a,Chr13b}. 

%
\begin{table*}[t!]
  \centering
  \caption{Two small example databases, \textbf{D}$_A$ and
    \textbf{D}$_B$, with missing values and variations, and where
    RecordID and EntityID represent record and entity identifiers,
    respectively.} \label{temp-table}
	\begin{small}
	\begin{tabular}{ccccccccc} \hline \noalign{\smallskip}
			Database~& RecordID &~EntityID~& ~FirstName~ &
			~LastName~ &
			BirthDate & StreetAddress & City & PhoneNumber\\
			\noalign{\smallskip} \hline\noalign{\smallskip}
			\multirow{3}{*}{\textbf{D}$_A$} &
			$r_1$ & $e_1$ &
			Peter &	Smith & 
			~1981-11-25~ & ~ & ~London~ & ~ \\
			~ & $r_2$ & $e_2$ &
			Peter & ~ & 
			1981-11-25 & ~43 Skye Pl~ & ~Dublin~ & +353 456 785\\
			
			~ & $r_3$ & $e_3$ & 
			Anne  &	Miller & 
			1991-09-11 & ~43 Skye Plc~ & ~ & +353 456 785\\
			\noalign{\smallskip} \hline \noalign{\smallskip}
			\multirow{2}{*}{\textbf{D}$_B$} &
			$r_4$ & $e_2$ & 
			Peter &	Smith & 
			1981-11-25 & 43 Skye Pl & ~ & ~\\
			~ & $r_5$ & $e_3$ &
			Ann & Myller & 
			~ &	43 Skye Place & ~Dublin~ & +353 456 785\\
			\noalign{\smallskip} \hline
		\end{tabular}
     \end{small}
\end{table*}


One major data quality aspect that so far has only seen limited
attention in RL is missing data~\cite{Ani19,Fer18,Gol15,Ong14}.
There are various reasons why missing values can occur, ranging
from equipment malfunction or data items not considered to be
important, to deletion of values due to inconsistencies, or even
the refusal of individuals to provide information for example when
answering surveys. Missing data can be categorised into 
different types~\cite{Lit20}.

Data \emph{missing completely at random} (MCAR) are missing values
that occur without any patterns or correlations at all with any
other values in the same record or database. For example, if a first
name is missing for an individual, then neither last
name, address, age, nor gender, can help predict the missing first
name value (assuming no external database is accessible where a
complete record of that person is available). With data
\emph{missing at random} (MAR) one can assume that a missing value
can be predicted by other values in the same record and/or database.
As an example, for a record of a surgeon in an employment database,
her salary could be predicted by averaging the salaries of all
other (female) surgeons in that database. Data \emph{missing not at
random} (MNAR) do occur for some specific reasons, for example if a
patient in a medical study suffered from a stroke and therefore did
not return to re-examinations then there will be missing values for
this patient in the study's database. Finally, \emph{structurally
missing} data are values that are missing because they should not
exist, and missing is their correct value. For example, young
children should not have an occupation.
Structurally missing data are different to MNAR data because for
the former the correct actual value is an empty value, while for
MNAR data there does exist a correct value, but not recorded.

Missing data can often occur in the QID attributes used to link 
databases. While data imputation can be applied with the aim to 
fill such missing values before the databases are 
linked~\cite{Her07,Lit20}, in our work we assume that
not all missing values have been imputed. Specifically, imputation
is not possible for missing values of types MCAR and structurally
missing. 
For example, as shown in Table~\ref{temp-table}, if the 
street address is missing for an individual (like for $r_1$), then 
no other QID value of this record can help to predict this missing
last name value. 
Missing values can lead to lower similarities between
records and thereby affect linkage quality. If alternatively those
records or QIDs with missing values are not used for a linkage at
all, then linkage quality will also likely suffer.

As an example, given the five records of three entities in the two
databases \textbf{D}$_A$ and \textbf{D}$_B$ 
in Table~\ref{temp-table}, a linkage between 
\textbf{D}$_A$ and \textbf{D}$_B$ would potentially classify the
record pairs ($r_1$, $r_4$), and ($r_2$, $r_4$) as matches because
they have similar QID values, and it would potentially classify
records $r_3$ and $r_5$ to refer to different entities because they
have a low similarity due to name variations and missing values.
Therefore, this linkage will be of poor quality due to the wrongly
matched record pair ($r_1$, $r_4$) and the non-matched true matching
record pair ($r_3$, $r_5$). 

%
\smallskip

\textbf{Contribution}:
We propose a novel RL technique 
that is applicable in situations where QID values contain errors
and variations, and importantly that can be missing. For each record
in a database to be linked, our approach first generates several
distinct \emph{attribute signatures} by concatenating values from a
set of QID attributes~\cite{Zha18}. We propose an
Apriori~\cite{Ceg06} based technique to select suitable
QID attributes for signature generation. As we detail in
Section~\ref{sec:methodology}, the aim of attribute signatures is
to uniquely describe records such that record pairs can be matched
based on the number of common attribute signatures they share.

We then utilise the relationships between entities to identify
matching records assuming they contain missing or dirty values in
their QID attributes. We first generate a graph of records based
on the relationships between them, for example the roles of
individuals in a census household (such as \emph{father of} or
\emph{spouse of}), or shared attribute values such as addresses or
phone numbers. We then generate a \emph{relational signature}
for each record based on its neighbourhood in this graph, and
calculate the similarities between relational signatures of pairs
of records.

We analyse our approach in terms of its computational complexity,
and evaluate linkage quality using large real-world data sets, 
including 
simulated German census databases containing in total over 
130 million records.
As the experimental results in Section~\ref{sec:exp} show, our 
approach scales to large databases and incorporating relational
signatures can help to substantially improve linkage quality 
compared to several baseline approaches~\cite{Ena19,Fel69,Ong14,Zha18}.



\section{Related Work}
\label{sec:relwork}

Missing data has always been a challenge for linking records across
databases because missing QID values can result in lower linkage 
quality. Ong et al.~\cite{Ong14} introduced three methods to improve 
the accuracy of probabilistic RL~\cite{Fel69} when records have
missing values: redistribute matching weights of attributes that
have missing values to non-missing attributes, estimate the
similarity between attributes when values are missing in a record,
or use a set of additional attributes to calculate similarities if
a primary QID value is missing.

Enamorado et al.~\cite{Ena19} proposed an approach based on the
probabilistic RL model~\cite{Fel69} that allows for missing data
and the inclusion of auxiliary information when linking records.
This approach uses the Expectation-Maximisation (EM)
algorithm~\cite{Dem77} to estimate matching weights in an
unsupervised manner~\cite{Win00}. An approach similar to~\cite{Ena19}
was proposed by Zhang et al.~\cite{Zha20} for adjusting match
weights of a record pair when missing data occurred using the EM
algorithm, but does not scale to large databases
due to its computationally expensive weight estimations.

Goldstein and Harron~\cite{Gol15} developed an approach to link
attributes of interest for a statistical model between a primary
database and one or more secondary databases that contain the
attributes of interest. The approach exploits relationships between
attributes that are shared across the databases, and treats the
linkage of these databases as a missing data problem. The approach
uses imputation to combine information from records in the
different databases that belong to the same individual, and
calculates match weights to correct selection bias.

Ferguson et al.~\cite{Fer18} proposed an approach to improve RL
when the databases to be linked contain missing values by using a
modification of the EM algorithm~\cite{Dem77} that considers 
both the imputation of missing values as well as correlations 
between QID attribute values. 
%
Aninya et al.~\cite{Ani19} analysed how different blocking
techniques are affected by missing values. 
They showed that techniques that insert each record
into multiple blocks, such as canopy clustering and suffix array
indexing~\cite{Chr12a}, performed best when the attributes used
for blocking contained missing values.

In contrast to these existing methods~\cite{Ani19,Fer18,Gol15,Ong14},
our approach does not consider imputation or estimation of
similarities when QID values are missing. Instead, we exploit
relational information between records to
increase the overall similarities between record pairs even if
their attribute similarities are low due to missing values,
errors, or variations. To the best of our knowledge this work is
the first to consider relational information for similarity
calculations in the linkage process when records 
contain missing values.  

The use of combinations of QID values (called \emph{linkage keys})
is commonly used for deterministic linkage~\cite{Chr12a}. A
single attribute combination (such as SLK-581~\cite{Kar10a}) or
multiple attribute combinations (such as p-signatures~\cite{Zha18}),
which can improve the likelihood of identifying
matches~\cite{Rand19}, can be used to generate linkage keys to
identify matches. However,
linkage keys cannot tolerate differences in QID values, nor can
they handle missing values or use additional information such as
relationships between records~\cite{Kar10a,Rand19,Zha18}.

\section{Problem Formulation}
\label{sec:problem}

Without loss of generality, we
assume two deduplicated databases~\cite{Chr12a}, \textbf{D}$_A$ and
\textbf{D}$_B$, such that each entity is represented by only one
record in a database. 
%
%
We define two disjoint sets, $\mathbf{M}$ and $\mathbf{U}$, from
the cross-product 
\textbf{D}$_A$×\textbf{D}$_B$. 
A record pair $(r_i,r_j)$, with $r_i \in \mathbf{D}_A$ and $r_j \in
\mathbf{D}_B$, is a member $\mathbf{M}$ (true matches) if
$(r_i,r_j)$ represents the same entity; otherwise $(r_i,r_j)$ is a
member of $\mathbf{U}$ (true non-matches) and $r_i$ and $r_j$
represent two different entities. With the true class ($\mathbf{M}$
or $\mathbf{U}$) unknown, the record linkage process attempts to
accurately classify each record pair as belonging to either
$\mathbf{M}$ or $\mathbf{U}$~\cite{Chr12a,Fel69}.


As we detail in Section~\ref{sec:methodology}, the aim of our
approach is to link records between databases based on their values
in QID attributes, as well as relationship information between
records. We first generate a set of \emph{attribute signatures} for
each record, where each signature is formed by concatenating one or
more tokens (substrings) extracted from QID values~\cite{Zha18}. 
These tokens are generated by applying various string manipulation 
functions. Any pair of records with the same values for a given 
set of QID attributes will have the same attribute signature, 
thus leading to the identification of matching records. We define
an attribute signature as: 

\begin{definition}[Attribute Signature]
Let $\mathbf{D}$ be a database where each record $r\in \mathbf{D}$
contains values $r.a$ in a set of QID attributes $a\in A$. Let 
$S=[a\in A :\ C(a)]$ be a list of attributes where each $a \in S$
is selected according to a certain criteria $C$. The attribute
signature of $r$ using $S$ is defined as $\mathcal{A}_S(r,S)$ =
concat$([g_a(r.a):a \in S])$, where $concat()$ is the string
concatenation function and $g_a()$ is a manipulation function
applied on the value $r.a$ of attribute $a$.
\end{definition}
%

For example, let us assume the attributes selected for attribute
signature generation are based on a criteria $C$ of an attribute 
having a maximum of 20\% of records in a database with missing
values. Thus, assuming an attribute combination  
$S_1=[$\textit{FirstName}, \textit{LastName},
$yearOf$(\textit{BirthDate}$)]$ is selected, the attribute signature
generated for record $r_1$ in Table~\ref{temp-table} is
\emph{PeterSmith1981}, where the values 1981 is a token and the
string manipulation function $yearOf()$ returns the year of a date
of birth value. 

No signature can be generated for $r$ if a QID value
used in $S$ is missing for $r$~\cite{Zha18}. 
Furthermore, errors and variations in QID values can result in
different signatures, which can lead to record pairs to be
classified as false matches or false non-matches. For example,
assuming the same attribute combination as $S_1$, the records
$r_1$ and $r_4$ in Table~\ref{temp-table} will be classified as a
match because they have the same signature $\mathcal{A}_S(r_1,S_1) =
\mathcal{A}_S(r_4,S_1)=$ \emph{PeterSmith1981}, while $r_3$ and $r_5$
will be classified as non-match since no attribute signature can be
generate for $r_5$ due to its missing date of birth value.

To improve similarities when records contain missing values, or
errors and variations, we use the relationships between records
in a database, 
as identified through some attributes such as family relationship,
address, or phone number. For example, the records $r_2$ and $r_3$ in 
Table~\ref{temp-table} can be considered as related because 
they share the same phone number.
We represent records and their relationships in a
\emph{record graph}, defined as: 

\begin{definition}[Record Graph]
Let $G = (V, E)$ be an undirected graph where $V$ is the set of
vertices, and each $v\in V$ represents a record $r \in \mathbf{D}$.
The set of edges of $G$ is defined as $E=\{(v_i,v_j) : v_i,v_j \in
V\ \wedge\ v_i \overset{a} {\longleftrightarrow}v_j\}$ where $v_i
\overset{a} {\longleftrightarrow} v_j$ represents a relationship
over an attribute $a \in A$ between records $r_i$ and $r_j$. 
\end{definition}


We then generate a \emph{relational signature} for each $v \in G$
based on its graph neighbourhood~\cite{Agg10}, such as the degree of
$v$, where we generate $m \ge 1$ features for the relational
signature of a record. We define a relational signature as:

\begin{definition}[Relational Signature]
The relational signature of a record $r \in \mathbf{D}$ based on
its corresponding vertex $v \in G$, using its neighbourhood
$N(v)=\{v_j : (v,v_j) \in G.E\}$ is defined as $\mathcal{R}_S(r)$
= $\langle f_1(N(v))$, $\cdots$, $f_m(N(v))\rangle$, with $f_i$
being a certain graph feature generated from the neighbourhood
$N(v)$.
\end{definition}

As we describe next, the features $f_i$ that can be used in
relational signatures include degree and density using the egonet of
a vertex~\cite{Agg10}, or we can use the QID values of neighbouring
vertices (records) as features for a given vertex.
%
For example, records $r_3$ and $r_5$ in Table~\ref{temp-table}
can be matched based on their respective neighbouring records $r_2$
and $r_4$, given they share a highly similar street address, and
generate the same attribute signature
$\mathcal{A}_S(r_2,S_2) = \mathcal{A}_S(r_4,S_2) =$
\emph{Peter1991Skye} for
$S_2=[$\textit{FirstName}, $yearOf$(\textit{BirthDate}),
$streetName($\textit{StreetAddress}$)]$.

%


\section{Signature based Record Linkage in the Presence of Missing
Data}
\label{sec:methodology}

Our approach consists of three main steps. We first
provide an adaptive method to select attribute combinations (from
the set of QID attributes $A$) that are suitable to generate attribute
signatures. We then 
generate several attribute signatures per record, as well as one relational
signature per record based on the record graph we build for each database 
to be linked.
Finally, 
we use the generated attribute and relational signatures to identify
matching records.



\subsection{Attribute Selection}

The choice of QID attributes plays an important role for attribute
signature generation. The aim of each attribute signature is to 
distinctively describe each record, such that if two records share
the same attribute signature it is highly likely that they belong
to the same entity. Due to errors, variations, and missing values
in QIDs, however, a certain attribute combination might not be
suitable as an attribute signature. For example, 
$S_3=[$\textit{FirstName}, \textit{BirthDate}$]$ in
Table~\ref{temp-table} would generate the same attribute
signature \emph{Peter1981-11-25} for records $r_1$, $r_2$, and
$r_4$, and result in the false matching of record pair ($r_1, r_4$).  

One way to overcome this issue is to select a set $S$ of
attributes based on domain knowledge~\cite{Zha18}. However, such 
knowledge is not always available, or it might be inadequate for a
given linkage project, leaving a user to randomly select combinations
of (potentially not suitable) QIDs for attribute signature generation.
%
We therefore propose a method to select suitable QID attributes for
signature generation based on their data characteristics. We
consider two aspects in the selection process, (1)
\emph{completeness}, which measures the amount of missing values in
each attribute, and (2) \emph{Gini impurity}~\cite{Han11}, which
measures the homogeneity of an attribute with respect to the number
of its distinct values.

\begin{figure}[!t]
\begin{center}
  \label{algo:attr_select}
  \begin{tabular}{ll} \hline \noalign{\smallskip}
  \multicolumn{2}{l}{\textbf{Algorithm~1: \emph{Attribute
    Selection}}} \\ 
  \noalign{\smallskip} \hline \noalign{\smallskip}
  \multicolumn{2}{l}{Input:} \\
  \multicolumn{2}{l}{- $\mathbf{D}$: \hspace{1mm}\,Database to be
      linked} \\
  \multicolumn{2}{l}{- $A$: \hspace{1mm} List of QID
    attributes} \\
  \multicolumn{2}{l}{- $n_a$: \hspace{1mm}Number of combinations required} \\    
  \multicolumn{2}{l}{- $\alpha$: \hspace{2mm}Score weight, with
    $0\le\alpha\le 1$} \\    
  \multicolumn{2}{l}{- $c_t$: \hspace{1.6mm}Attribute selection threshold,
    with $0\le c_t\le 1$} \\
    \noalign{\smallskip}
  \multicolumn{2}{l}{Output:} \\
  \multicolumn{2}{l}{- $\mathbf{C}_S$: List of candidate attribute
    combinations} \\ \noalign{\smallskip}
  1:  & $\mathbf{C} = \{\}$, $\mathbf{T} = [\ ]$, $k = 2$
        \hspace*{\fill} // \small Initialise variables \\
  2:  & $\mathbf{C}[1] = \mathit{genCombinations}(A)$
        \hspace*{\fill} // \small Set initial combinations \\
  3:  & \textbf{while} $ k \le |A|$ \textbf{do:} 
        \hspace*{\fill} // \small Loop over all sizes of combinations \\
  4:  & \hspace{2mm} $\mathbf{C}[k] =
        \mathit{genCombinations}(\mathbf{C}[k-1])$  \\  
  5:  & \hspace{2mm} \textbf{foreach} $S \in \mathbf{C}[k]$
        \textbf{do:} 
        \hspace*{\fill} // \small Loop over each combination $S$\\
  6:  & \hspace{5mm} $s_c = \mathit{getCompletenessScore}(S,
        \mathbf{D})$ 
		\hspace*{\fill} // \small Completeness of $S$ \\
  7:  & \hspace{5mm} $s_g = \mathit{getGiniScore}(S, \mathbf{D})$ 
		\hspace*{\fill} // \small Gini impurity of $S$ \\
  8:  & \hspace{5mm} $s = \alpha\cdot s_c + (1-\alpha)\cdot s_g$
        \hspace*{\fill} // \small Calculate overall score of $S$ \\
  9:  & \hspace{5mm} \textbf{if} $s \ge c_t$ \textbf{do:}
        \hspace*{\fill} // \small Check if score $s$ is at least $c_t$ \\
  10: & \hspace{8mm} $\mathbf{T}$.add($(S, s)$)
        \hspace*{\fill} // \small Add the combination and its score
        \\        
  11: & \hspace{8mm} $\mathbf{T}$ =
        $\mathit{removeSubCombinations(S, \mathbf{T})}$
        \hspace*{\fill} // \small Remove subsets\\
  12: & \hspace{5mm} \textbf{else:} 
        \hspace*{\fill} // \small Score $s$ is less than $c_t$\\
  13: & \hspace{8mm} $\mathbf{C}[k]$.remove($S$)
        \hspace*{\fill} // \small Remove the combination $S$\\
  14: & \hspace{2mm} $k= k + 1$
        \hspace*{\fill} // \small Increment $k$\\        
  15: & $\mathbf{C}_S = getTopCombinations(\mathbf{T},
        n_a)$ 
		\hspace*{\fill} // \small Get best combinations \\
  16: & return $\mathbf{C}_S$ \\
      \hline
  \end{tabular}
\end{center}
\end{figure}

As we outline in Algorithm~1, we follow an Apriori based 
approach~\cite{Agr93,Ceg06} 
where we use an iterative search over attribute combinations of
length $k$ generated from suitable attribute combinations of length
$k-1$. First, in lines 1 and 2, we initialise an inverted index
$\mathbf{C}$ to store the combinations we need to process. In lines
3 to 13 we then loop over each attribute combination of size $k$, with
$2 \le k \le |A|$. 
In line 4, using the function
$\mathit{genCombinations()}$ we generate the set $\mathbf{C}[k]$
of attribute combinations of size $k$ based on the previously
generated combinations $\mathbf{C}[k-1]$ of size $k-1$. 

Following the Apriori principle~\cite{Agr93}, to generate an 
attribute combination $S$ of size $k$ requires that all its attribute
combinations of size $k-1$ are in $\mathbf{C}[k-1]$, which means
all its subsets must satisfy the selection criteria we detail below.
For example, for the attribute combination $[a_1,a_2,a_3]$ of size
$k=3$, where $a_i \in A$, all three combinations $[a_1,a_2]$,
$[a_2,a_3]$, and $[a_1, a_3]$ must be in $\mathbf{C}[2]$.

In lines 5 to 7, the algorithm iterates over each combination $S \in
\mathbf{C}[k]$ and calculates the completeness and Gini impurity
scores $s_c$ and $s_g$, respectively. We calculate $s_c =
(|\mathbf{D}| - m_S)/|\mathbf{D}|$, where $|\mathbf{D}|$ is the
number of records in $\mathbf{D}$ and $m_S$ is the number of 
records with missing values for any attribute in
$S$. We calculate the Gini impurity of $S$
as $s_g = \sum_{v \in V} P(v) \cdot (1-P(v))$, where $V$ is the
list of unique values in 
$S$ and the probability $P(v)=f_v/|\mathbf{D}|$ with $f_v$ being 
the frequency of 
value $v\in\mathbf{D}$. 
In lines 8 and 9, a combination $S$ is selected as a candidate
attribute signature if the weighted score $s = \alpha \cdot s_c +
(1-\alpha)\cdot s_g$ is at least the threshold $c_t$, where $\alpha$
determines the importance of completeness over Gini impurity.

If the score $s$ is at least $c_t$, in lines 10 and 11 we add $S$
along with its score $s$ to the list $\mathbf{T}$, and remove any
subsets of $S$ from $\mathbf{T}$ that were added in previous
iterations. This subset pruning prevents redundant attribute
combinations and reduces the costs of generating not required
attribute signatures. For instance, if the attribute combination
[\textit{FirstName}, \textit{LastName},
\textit{BirthDate}] is selected, then all its subsets
([\textit{FirstName}, \textit{LastName}],
[\textit{FirstName}, \textit{BirthDate}], and
[\textit{LastName}, \textit{BirthDate}]) are pruned because they
cannot provide more distinctive attribute signatures.

In line 13, we remove the combination $S$ from the set of 
combinations $\mathbf{C}[k]$ if its score $s$ is less than $c_t$. 
This ensures the supersets of $S$ will not be processed in 
subsequent iterations. For instance, if the score of the attribute
combination [\textit{FirstName}, \textit{LastName}] is less than
the threshold $c_t$, then it is unlikely that its superset, such
as [\textit{FirstName}, \textit{LastName}, \textit{BirthDate}], 
can provide more distinctive attribute signatures.

In line 14 we increment the value of $k$ by 1 allowing the 
algorithm to continue processing attribute combinations 
until $k=|A|$. Finally, in line 15 we select the $n_a$ 
attribute combinations with the highest scores in the list 
$\mathbf{T}$ to create the final list of combinations 
$\mathbf{C}_S$. As we show in Section~\ref{sec:exp}, 
increasing the number of attribute combinations, $n_a$, can 
lead to an increase of the overall runtime of our approach
as more signatures are generated per record, but potentially 
can also increase the linkage quality as more attribute 
signatures will be considered when comparing a pair of records. 



The complexity of Algorithm~1 is $O(|\mathbf{D}| \cdot
2^{|A|})$. 
The attribute selection threshold, $c_t$, can be used 
to increase or decrease the number of candidate attribute 
combinations that are pruned in each iteration, with lower 
values for $c_t$ generating more attribute combinations



\subsection{Signature Generation}

\begin{figure}[!t]
\begin{center}
  \label{algo:sig_generation}
  \begin{tabular}{ll} \hline \noalign{\smallskip}
  \multicolumn{2}{l}{\textbf{Algorithm~2: \emph{Signature
    Generation}}} \\ 
  \noalign{\smallskip} \hline \noalign{\smallskip}
  \multicolumn{2}{l}{Input:} \\
  \multicolumn{2}{l}{- $\mathbf{D}$: \hspace{1mm}\,Database to be
      linked } \\
  \multicolumn{2}{l}{- $A$: \hspace{2mm}List of QID attributes} \\      
  \multicolumn{2}{l}{- $\mathbf{C}_S$: \,List of candidate
    attribute combinations } \\
  \multicolumn{2}{l}{- $F$: \hspace{2mm}\,List of graph feature functions} \\    
  \multicolumn{2}{l}{- $p_t$: \hspace{1mm}\,Probability threshold, with $0\le p_t\le 1$} \\
  \noalign{\smallskip}
  \multicolumn{2}{l}{Output:} \\
  \multicolumn{2}{l}{- $\mathbf{S}$: Signature database} \\ \noalign{\smallskip}
  1:  & $\mathbf{S} = \{\}$, $\mathbf{A} = \{\}$,
        $L =\emptyset$ 
        \hspace*{\fill} // \small Initialise variables \\
  2:  & \textbf{foreach} $r \in \mathbf{D}$ \textbf{do:} 
        \hspace*{\fill} // \small Loop over each record in the database \\
  3:  & \hspace{2mm} $\mathbf{A}[r.id] = \emptyset$
        \hspace*{\fill} // \small Initialise to an empty set \\
  4:  & \hspace{2mm} \textbf{foreach} $S\in \mathbf{C}_S$
        \textbf{do:} \hspace*{\fill} // \small Loop over each combination \\
  5:  & \hspace{4mm} $\mathcal{A}_S =
        \mathit{genAttrSignature}(r, S)$
        \hspace*{\fill} // \small Attr. signature (Def.~3.1) \\
  6:  & \hspace{4mm} $\mathbf{A}[r.id]$.add($\mathcal{A}_S$) 
        \hspace*{\fill} // \small Add the generated signature to
        $\mathbf{A}$ \\
  7:  & \hspace{4mm} $L$.add($\mathcal{A}_S$) 
        \hspace*{\fill} // \small Add the generated signature to the set $L$\\
  8:  & \textbf{foreach} $\mathcal{A}_S \in L$ \textbf{do:} 
        \hspace*{\fill} // \small Loop over attribute signatures \\
  9: & \hspace{2mm} $p = \mathit{getSigProbability}(\mathcal{A}_S,
       \mathbf{A})$ \hspace*{\fill} // \small Calculate the probability \\
  10: & \hspace{2mm} \textbf{if} $p < p_t$ \textbf{do:}
        \hspace*{\fill} // \small Check if the probability is less than $p_t$ \\
  11: & \hspace{4mm} $\mathbf{A} =
        \mathit{removeSignature}(\mathcal{A}_S, \mathbf{A})$
        \hspace*{\fill} // \small Remove the signature \\
  12:  & $G = \mathit{genRecordGraph}(\mathbf{D}, A)$
        \hspace*{\fill} // \small Generate the graph (Def.~3.2) \\
  13: & \textbf{foreach} $r \in G$ \textbf{do:}
        \hspace*{\fill} // \small Loop over each record in $G$ \\
  14: & \hspace{2mm} $\mathcal{R}_S =
        \mathit{genRelSignature}(r, G, F)$
        \hspace*{\fill} // \small Rel. Signature (Def.~3.3) \\        
  15: & \hspace{2mm} $\mathbf{S}[r.id] = (\mathbf{A}[r.id],
        \mathcal{R}_S)$ \hspace*{\fill} // \small Add signatures of $r$ to $\mathbf{S}$ \\
  16: & return $\mathbf{S}$ \\ \hline
  \end{tabular}
\end{center}
\end{figure}

As shown in Algorithm~2, in the second step we
generate attribute and relational signatures for each record $r$ in
the database $\mathbf{D}$. In line 2 we loop over each $r \in
\mathbf{D}$, and initialise an inverted index list in $\mathbf{A}$
by adding the corresponding record identifier $r.id$ as a key and
with an empty set as values (line 3). Next, for each attribute
combination $S$ in the list of attribute combinations $\mathbf{C}_S$
from Algorithm~1, we generate the corresponding attribute signature
$\mathcal{A}_S$ (as per Definition~3.1) and add it to the sets
$\mathbf{A}[r.id]$ and $L$ (lines 5 to 7). Attribute signatures are
generated only if all attributes in $S$ have a value for a given $r$.

As we discussed in Section~\ref{sec:problem}, we require signatures
to be unique to a record because otherwise they can lead to
incorrectly linked records. The probability for a given attribute 
signature $\mathcal{A}_S$ to occur repeatedly in $n$ records
is governed by a Binomial distribution~\cite{Lit20}. 

The probability
of an attribute value combination in a record to become an attribute 
signature decreases if the value combination occurs in multiple 
records. We followed the same probability calculation described by 
Zhang et al.~\cite{Zha18} to calculate the probability $p$ of an
attribute signature $\mathcal{A}_S\in L$ to be considered as a
signature, which can be calculated as $p = P(\mathcal{A}_S\ |\ n =
|\{r.id : \forall_{r.id \in \mathbf{A}} \mathcal{A}_S \in
\mathbf{A}[r.id]\}|)=1/(1+\lambda^n\mu)$, with $0< \mu < 1 <
\lambda$. The two parameters $\lambda$ and $\mu$ control how fast
the maximum of $p$ decreases as the number of records $n$ that
have signature $\mathcal{A}_S$ increases~\cite{Zha18}.
We refer the reader to \cite{Zha18} for more details on 
this probability calculation. 

We keep $\mathcal{A}_S$ if its probability $p$ is at least the
threshold $p_t$ (line 10), otherwise we remove $\mathcal{A}_S$ from
all records in $\mathbf{A}$ in line 11. In line 12 we then generate
the record graph $G$ following Definition~3.2. In line 13 we loop
over each record $r \in G$ and in line 14 we generate the relational 
signature $\mathcal{R}_S$ for $r$ based on its local neighbourhood
structure in $G$. We provide a list $F$ of graph feature functions
to the function $genRelationalSignature()$ which generates a 
$\mathcal{R}_S$. 

To generate a relational signature we use different features that
describe the local graph neighbourhood of a vertex in $G$. Features
include degree and egonet density of a vertex~\cite{Agg10}, as well
as the attribute signatures of its neighbouring vertices for a
given vertex.
%
%
%
%
Finally, in line 15, we add the set of generated attribute signatures
$\mathbf{A}[r.id]$ and $\mathcal{R}_S$ of each record $r \in
\mathbf{D}$ into the signature database $\mathbf{S}$ using the
corresponding record identifier $r.id$ as the key. 

Generating the graph $G$ is of 
$O(|\mathbf{D}| + |\mathbf{D}|\cdot d/2)$, assuming the average
number of neighbours of a record in $G$ (its degree) is $d$. 
If we assume $|\mathbf{C}_S|$ attribute combinations have been 
generated by Algorithm~1, and each
record in $\mathbf{D}$ is on average represented by $l$ attribute
signatures, and $|F|$ is the number of features considered
to generate a relational signature, then the complexity of the 
signature generation process in Algorithm~2 is 
$O(|\mathbf{D}| \cdot(|\mathbf{C}_S| + l + d\cdot |F|))$.


\subsection{Signature based Record Matching}

\begin{figure}[!t]
\begin{center}
  \label{algo:sig_matching}
  \begin{tabular}{ll} \hline \noalign{\smallskip}
  \multicolumn{2}{l}{\textbf{Algorithm~3: \emph{Signature based
  Record Matching}}} \\ 
  \noalign{\smallskip} \hline \noalign{\smallskip}
  \multicolumn{2}{l}{Input:} \\
  \multicolumn{2}{l}{- $\mathbf{S}_A$: \hspace{6mm}\,Signature database
  of \textbf{D}$_A$} \\
  \multicolumn{2}{l}{- $\mathbf{S}_B$: \hspace{6mm}\,Signature database
  of \textbf{D}$_B$} \\
  \multicolumn{2}{l}{- $sim_A()$: \,\,Similarity function
    for attribute signatures} \\
  \multicolumn{2}{l}{- $sim_R()$: \,\,Similarity function
    for relational signatures} \\    
  \multicolumn{2}{l}{- $\beta$: \hspace{7mm}\,\,Similarity weight, with
     $0\le\beta\le 1$}\\
  \multicolumn{2}{l}{- $s_t$: \hspace{7mm}\,Similarity threshold, with
    $0\le s_t\le 1$}\\    
  \noalign{\smallskip}
  \multicolumn{2}{l}{Output:} \\
  \multicolumn{2}{l}{- $\mathbf{M}$: Matching record pairs} \\
  \noalign{\smallskip}
  1:  & $\mathbf{M} = \emptyset$
        \hspace*{\fill} // \small Initialise a set to hold matching record
        pairs \\
  2:  & $\mathbf{R} = \mathit{genRecordPairs}(\mathbf{S}_A,
        \mathbf{S}_B)$ \hspace*{\fill} // \small Generate record pairs \\
  3:  & \textbf{foreach} $(r_A.id, r_B.id)\in \mathbf{R}$ \textbf{do:} 
        \hspace*{\fill} // \small Loop over all record pairs \\ 
  4:  & \hspace{2mm} $\Lambda_A, \Lambda_B, \mathcal{R}_A,
        \mathcal{R}_B = \mathit{getSignatures}(\mathbf{S}_A[r_A.id],
        \mathbf{S}_B[r_B.id])$ \\ 
  5:  & \hspace{2mm} $s_A = \mathit{sim_A}(\Lambda_A, \Lambda_B)$
        \hspace*{\fill} // \small Calculate attr. signature
        similarity \\
  6: & \hspace{2mm} \textbf{if} $s_A\ge s_t$ \textbf{do:} 
        \hspace*{\fill} // \small Check if the similarity is at least $s_t$ \\
  7:  & \hspace{4mm} $\mathbf{M}$.add($(r_A.id, r_B.id)$) 
        \hspace*{\fill} // \small Add the record pair to $\mathbf{M}$\\
  8: & \hspace{2mm} \textbf{else:} \\        
  9:  & \hspace{4mm} $s_R = \mathit{sim_R}(\mathcal{R}_A, \mathcal{R}_B)$
        \hspace*{\fill} // \small Calculate rel. signature
        similarity \\        
  10: & \hspace{4mm} $s = \beta\cdot s_A + (1-\beta)\cdot s_R$ 
        \hspace*{\fill} // \small Calculate the similarity \\
  11: & \hspace{4mm} \textbf{if} $s\ge s_t$ \textbf{do:} 
        \hspace*{\fill} // \small Check if the similarity is at least $s_t$\\   
  12:  & \hspace{6mm} $\mathbf{M}$.add($(r_A.id, r_B.id)$) 
        \hspace*{\fill} // \small Add the record pair to $\mathbf{M}$ \\
  13: & return $\mathbf{M}$ \\
      \hline
  \end{tabular}
\end{center}
\end{figure}

In the last step, as outlined in Algorithm~3, we compare the
attribute and relational signatures generated for each record to
find matching record pairs that correspond to the same entity. We
first calculate the similarities between attribute signatures, and
if these are low due to missing QID values, or errors and
variations in QID values, we calculate the similarities between relational
signatures.

Our signature based matching process uses the two signature
databases $\mathbf{S}_A$ and $\mathbf{S}_B$ of databases
$\mathbf{D}_A$ and $\mathbf{D}_B$, respectively, as generated by
Algorithm~2. 
We use the function $sim_A()$ to calculate a similarity between
records based on the number of attribute signatures that they have
in common, while $sim_R()$ calculates similarities between the
features in relational signatures. For example, $sim_R()$ can be a
set based similarity function such as Jaccard or Dice similarity~\cite{Chr12a} 
if we use the QID values of neighbouring records as relationship features.

In line 2, the function $genRecordPairs()$ returns the set of all
record pairs $\mathbf{R} = \{ (r_A, r_B) :
\mathcal{A}_S\in\mathbf{S}_A[r_A.id],
\mathcal{A}_S\in\mathbf{S}_B[r_B.id]\}$ that have at least one
attribute signature in common.
%
In line 3, we loop over each $(r_A, r_B)\in\mathbf{R}$ and get its
respective set of attribute and relational signatures, $\Lambda_A,
\Lambda_B, \mathcal{R}_A$, and $\mathcal{R}_B$ in line 4. We
calculate the similarity $s_A$ between the sets of attribute
signatures $\Lambda_A$ and $\Lambda_B$, and if $s_A$ is at least a
user specified threshold $s_t$ then we add $(r_A, r_B)$ into the set
$\mathbf{M}$ of matching record pairs (lines 5 to 7).

For non-matching record pairs, in line 9 we calculate the
similarity $s_R$ between their relational signatures $\mathcal{R}_A$
and $\mathcal{R}_B$, and then calculate the weighted average
similarity $s$ using both $s_A$ and $s_R$. If $s$ is at least
$s_t$ then in line 12 the record pair $(r_A, r_B)$ is classified 
as a match and added to $\mathbf{M}$, which is returned in line 13.


The complexity of Algorithm~3 is $O(n)$ if we assume 
$n=|\mathbf{D}_A|=|\mathbf{D}_B|$ and each record in $\mathbf{D}_A$ 
and $\mathbf{D}_B$ has generated a unique attribute signature in the
signature generation step in Algorithm~2.

\section{Experimental Evaluation}
\label{sec:exp}

\begin{table*}[t!]
\centering
\caption{A summary of the number of records, percentage of records
  containing missing values, and the number of ground truth record
  pairs for each data set pair used in the experimental evaluation.}
\label{tab:dataset}
\begin{small}
\begin{tabular}{lccccccccc} \hline \noalign{\smallskip}
  ~ & \multicolumn{2}{c}{ DBLP - Scholar \hspace{2mm}} &
  \multicolumn{2}{c}{\hspace{1mm} DBLP - ACM \hspace{2mm}} &
  \hspace{2mm} UKCD \hspace{1mm} &
  \hspace{1mm} NCVR-20 \hspace{2mm} &
    NCVR-50 \hspace{2mm} & \multicolumn{2}{c}{NCVR-2017 - NCVR-2019} \\
   \noalign{\smallskip} \hline \noalign{\smallskip}
  Records & 2,616 & 64,263 & 2,616 & 2,294 & 18,699 & 877,012 &
    877,012 & 7,846,174 & 7,688,308\\ 
  Missing \% & 17\% & 54\% & 20\% & 20\% & 24\% & 20\% & 50\% & 12\% & 10\%\\
  Ground truth pairs~~ & \multicolumn{2}{c}{2,386} &
    \multicolumn{2}{c}{2,220} & 18,354 & 877,012 & 877,012 & \multicolumn{2}{c}{6,948,248}\\
  \noalign{\smallskip} \hline
\end{tabular}%
\end{small} 
\end{table*}

We use six real-world data sets to empirically evaluate our
proposed approach, as summarised in Table~\ref{tab:dataset},
where each data set contains a pair of databases. Two are
bibliographic data sets~\cite{Kop10}, DBLP-Scholar and DBLP-ACM,
where their entities are academic publications, and each record
contains title, authors, venue, and year of publication. We added
missing values randomly, following the missing completely at
random (MCAR) missing data category, to authors, venue, and year
of publication, ensuring each record contains at most two missing
values. The relationships between publications are identified based
on their common authors.

\begin{table*}[t!]
\centering
\caption{Average runtime results (in seconds) for linking the
  different data set pairs.}
\label{tab:runtime}
\begin{small}
\begin{tabular}{lccccccc} \hline \noalign{\smallskip}
  \multicolumn{2}{l}{} & ~~DBLP - Scholar~~ & 
    DBLP - ACM & ~~~UKCD~~~ & ~~~NCVR-20~~~ &
    ~~NCVR-50 & NCVR-2017 - NCVR-2019\\
    \hline\noalign{\smallskip}
  \multicolumn{2}{l}{ProbLink~\cite{Fel69}}         & 143 & 12 & 164
    & 1,155 & 1,143 & 38,468 \\ 
  \multicolumn{2}{l}{ReDistLink~\cite{Ong14}} & 141 & 11 & 148 &
    1,090 & 1,125 & 34,766 \\ 
  \multicolumn{2}{l}{P-Sig~\cite{Zha18}}      & 9 & 6 & 8 & 45 & 47 & 892
    \\ 
  \multicolumn{2}{l}{FastLink~\cite{Ena19}}     & 15 & 26 & 38 & 267 & 279 & 6,482
    \\ 
  \multicolumn{2}{l}{\textbf{RA-Sig} (steps 1\,/\,2\,/\,3)} & 5 / 1 / 5 & 5 / 1 / 2 & 10 / 2 / 3 & 
  112 / 43 / 17 & 123 / 39 / 19 & 861 / 363 / 198 \\ \noalign{\smallskip} \hline

\end{tabular}%
\end{small}
\end{table*}

The UKCD data set consists of census records of 12,044 households
collected from 1851 to 1901 in 10 year intervals for the town of
Rawtenstall and surrounds in the United Kingdom~\cite{Fu14a}. We
use the role of individuals in a household to identify the
relationships between records. We used first name, middle name,
surname, address, occupation, gender, age, and birth parish as
QIDs, where 24\% of records contain missing values in at most four
of these attributes. 

We assume the missing data in the UKCD data set represents the MCAR,
MAR, and MNAR categories. This is because the missing values in
attributes such as first or middle name cannot not be imputed using
other QID values, which makes them MCAR data. The missing values in
address could be imputed based on the other available information
of the household as we assume all members lived at the same address
when the census was collected, hence they follow the MAR category.
Finally, some missing values in the occupation attribute follow 
MNAR as records about children do not contain an occupation. 

The NCVR data sets contain records from an October 2019
snapshot of the North Carolina voter registration database
(available from: \url{http://dl.ncsbe.gov/}). We extracted 309,877
household units from this snapshot and created two subsets, NCVR-20
and NCVR-50, where we randomly created (assuming MCAR) at most 
five missing values in the QID attributes first, middle, and 
last names, street address, city, and zip code in 20\% and 50\% of
records, respectively. To evaluate our approach with data of 
different quality, we then applied various corruption functions
using the GeCo data corruptor~\cite{Chr13b} on between 1 to 3 
randomly selected QID values of 20\% of records. For relationship 
information we considered if two records shared the same 
last name and address. 

Finally, we use two full NCVR snapshots from October 2017 and
October 2019, named as NCVR-2017 and NCVR-2019, respectively, as 
two databases to be linked. We use the same QID attributes first name,
middle name, last name, street address, city, and zip code as the 
other NCVR data sets for linkage. The last name and address QIDs
did not contain any missing values, and we therefore consider two
voters as related if they share the same address and last name.
Among the 89\% voters who appear in both snapshots, 1\%, 3\%, 1.3\%,
17\%, 10\%, and 12.5\% of voters have changed their first name, 
middle name, last name, address, city, and zip code, respectively. 

We compare our approach (\textbf{RA-Sig}) with four baseline linkage
techniques. We use traditional probabilistic record linkage
(\emph{ProbLink}) as proposed by Fellegi and Sunter~\cite{Fel69} as
a baseline given this approach is widely used in practical record
linkage applications~\cite{Fer18,Gol15}.
%
We also use the method by Ong et al.~\cite{Ong14} (\emph{ReDistLink})
that redistributes weights of QID attributes that have missing values
to non-missing QID attributes, where these attribute weights are 
calculated based on the probabilistic RL method~\cite{Fel69}. 
%

The third baseline is the probabilistic signature approach
(\emph{P-Sig}) proposed by Zhang et al.~\cite{Zha18}, which
identifies attribute signatures that distinctively describe
records, without however considering errors, variations, nor
missing values.
%
Finally, we use the recently proposed probabilistic linkage approach
by Enamorado et al.~\cite{Ena19}, named \emph{FastLink}. This approach
also utilises the Expectation Maximisation algorithm based 
probabilistic RL model~\cite{Dem77,Fel69} which estimates matching 
weights considering missing values and auxiliary information when 
linking records. 


Based on a set of parameter sensitivity experiments, in Algorithm~1
we set $n_a=[1, 5, 10]$, $\alpha=0.5$, and $c_t=[0.5, 0.7, 0.9]$.
Following~\cite{Zha18}, we set $\lambda=1.2$, $\mu=0.2$, and
$p_t=[0.5, 0.7, 0.9]$ in Algorithm~2.
We use the attribute signatures of neighbouring vertices
as features in Algorithm~3.
The Jaccard coefficient~\cite{Chr12a} is used for both
$sim_A()$ and $sim_R()$, and in Algorithm~3 we set $\beta=0.5$.
Following~\cite{Ong14} we set the similarity threshold $s_t$ in
Algorithm~3 to range from 0.5 to 1.0 (in 0.1 steps) for all 
techniques. 

We evaluated scalability using runtime, and linkage quality using
precision and recall~\cite{Chr12a}. Precision measures the number 
of true matched record pairs against the total number of record
pairs generated by a particular approach; and recall measures the
number of true matched record pairs against the total number of
record pairs in the linked ground truth data~\cite{Han18}. We do not
use the F-measure for evaluation because recent research has shown
that F-measure is not suitable for measuring linkage quality in
record linkage due to the relative importance given to precision
and recall which  depends upon the number of predicted
matches~\cite{Han18}.

\begin{figure*}[!t]
  \centering
  \includegraphics[width=0.32\textwidth]
  {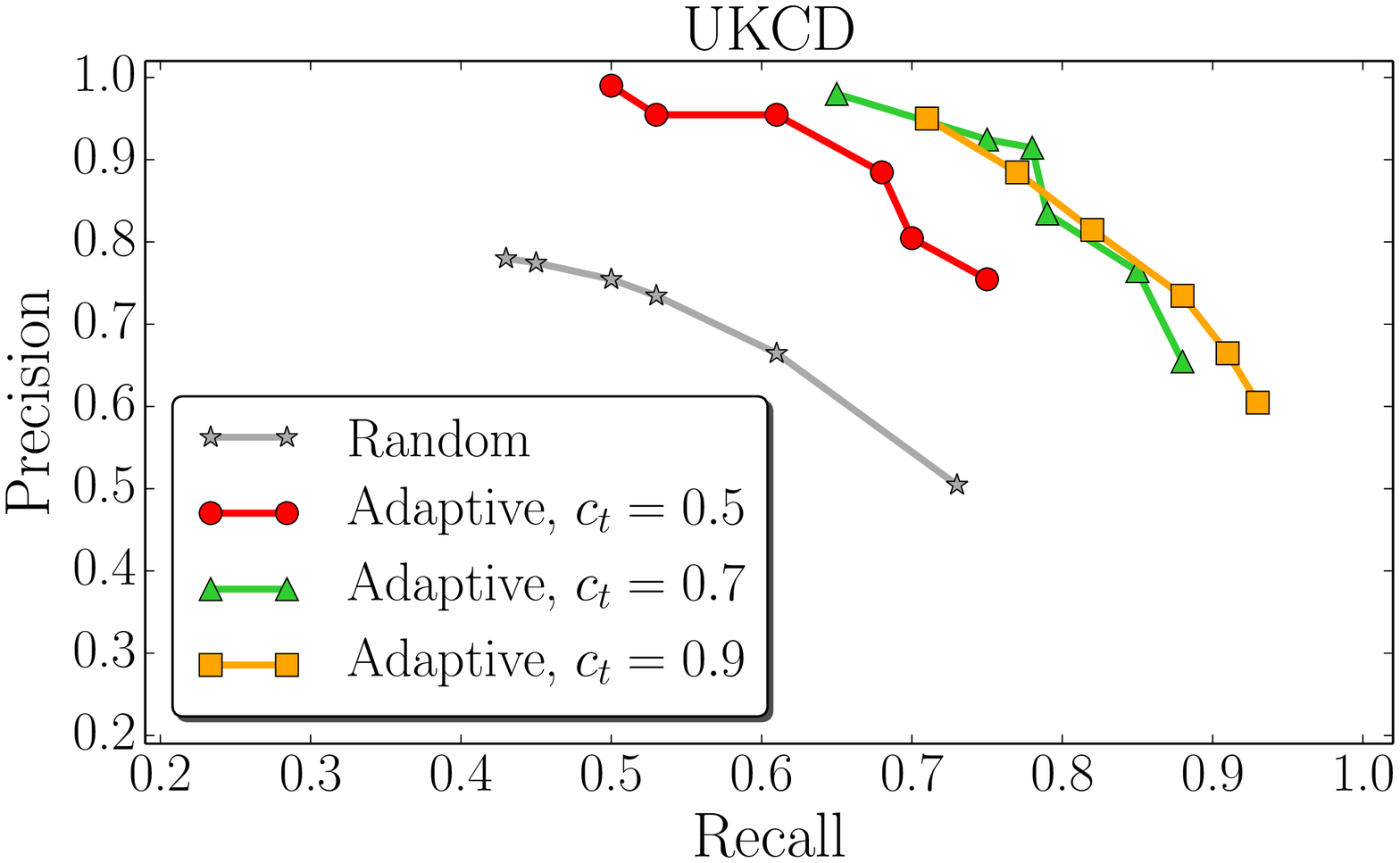}
  \includegraphics[width=0.32\textwidth]
  {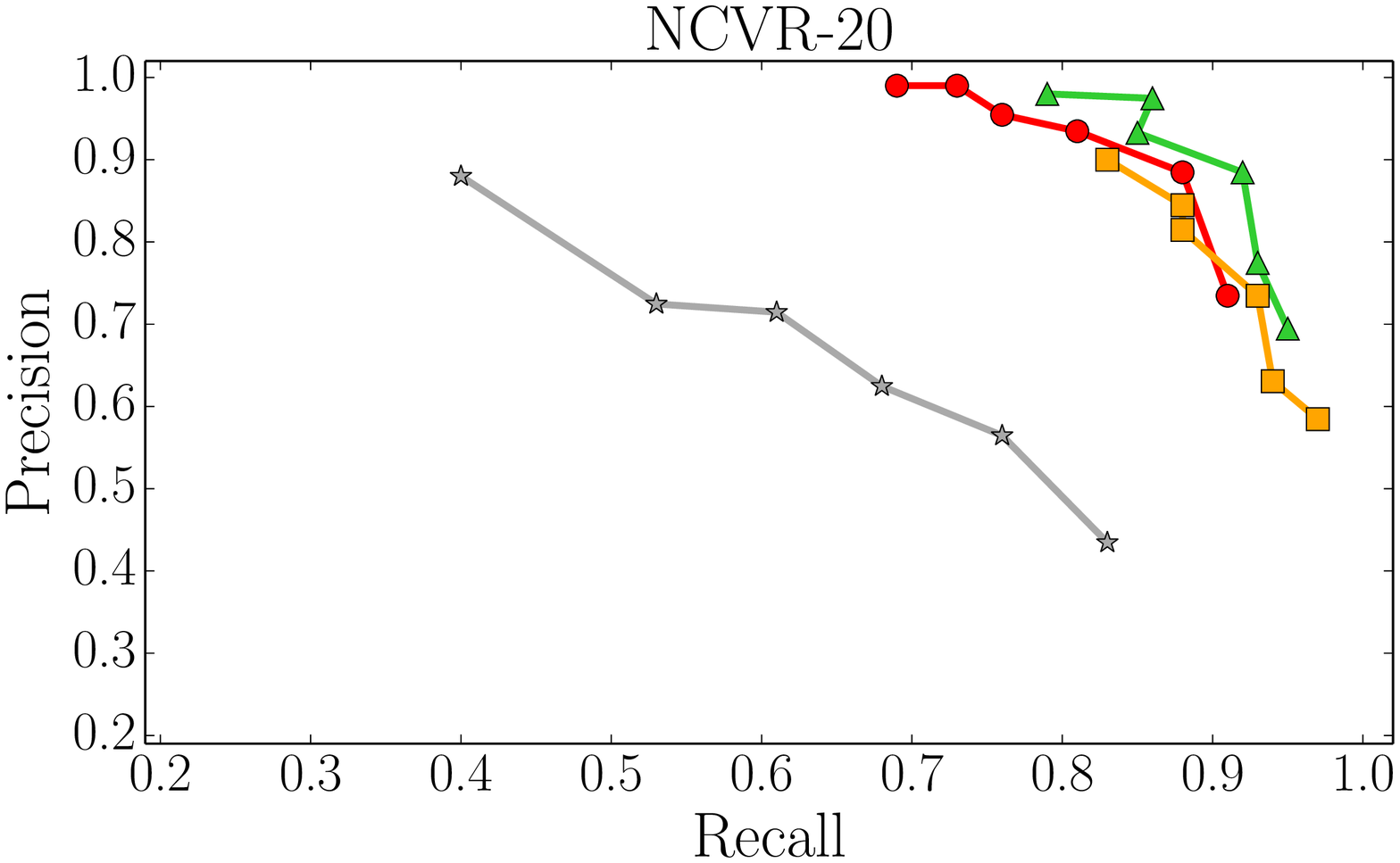}
  \includegraphics[width=0.32\textwidth]
  {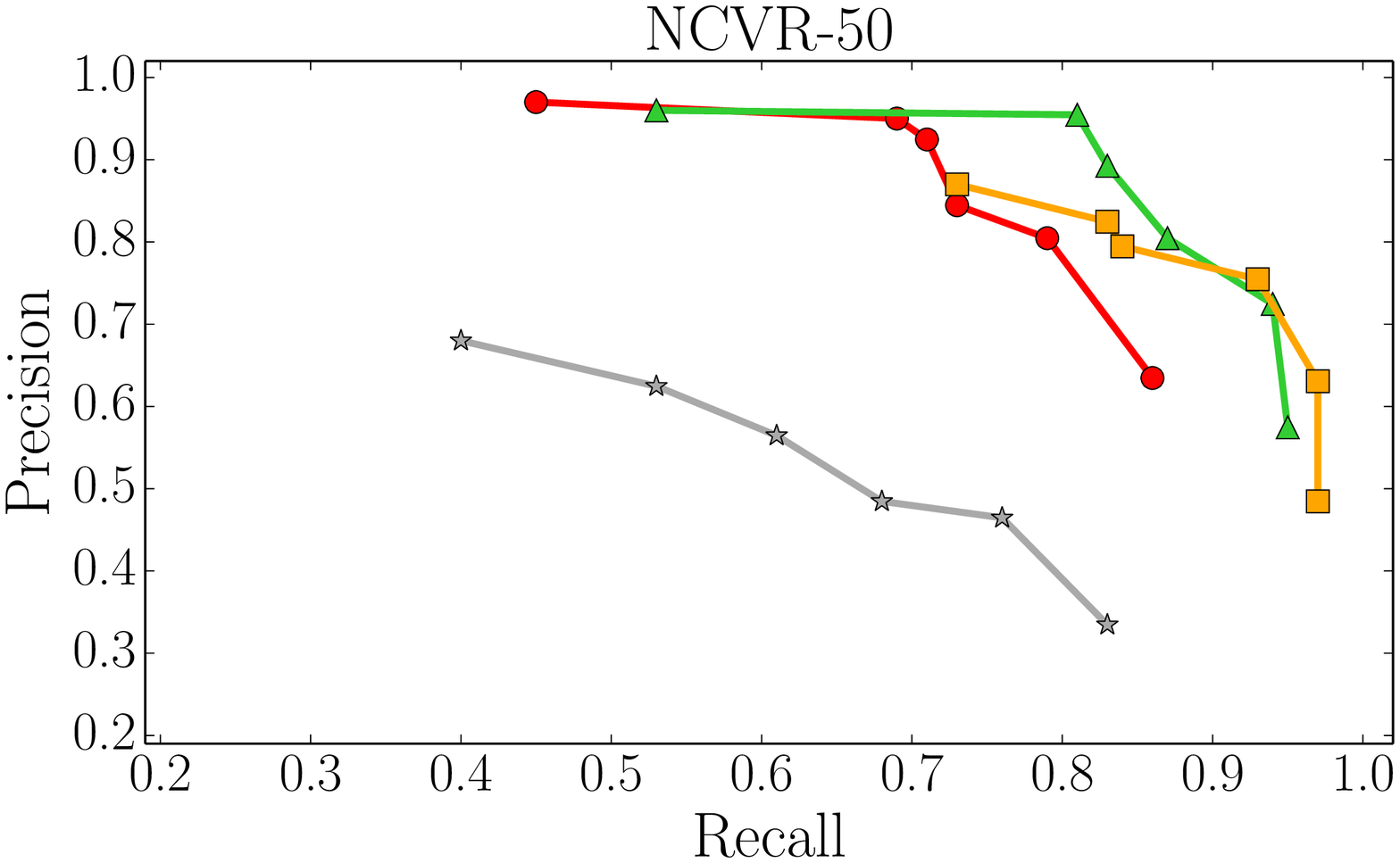}
  \includegraphics[width=0.32\textwidth]
  {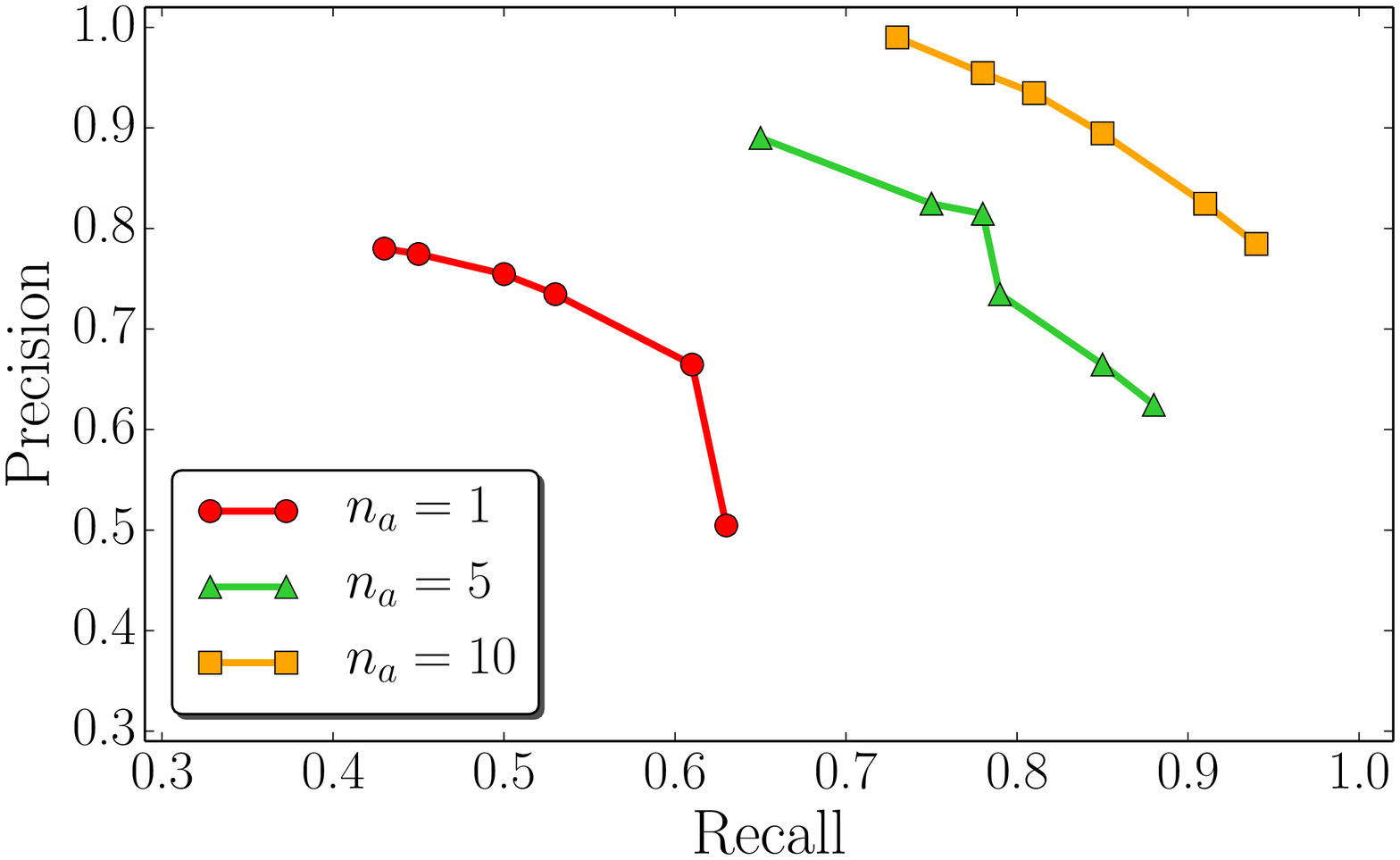}
  \includegraphics[width=0.32\textwidth]
  {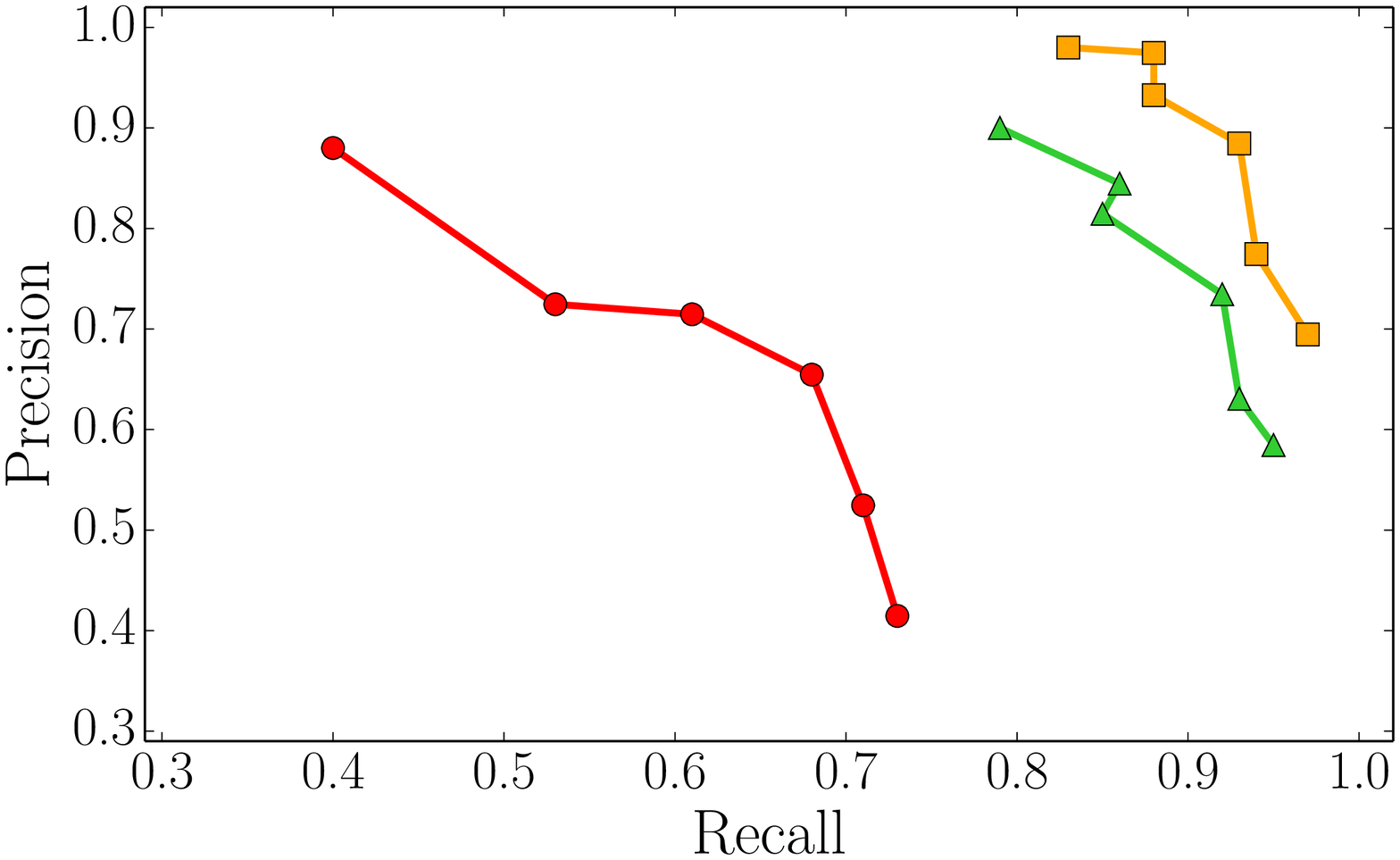}
  \includegraphics[width=0.32\textwidth]
  {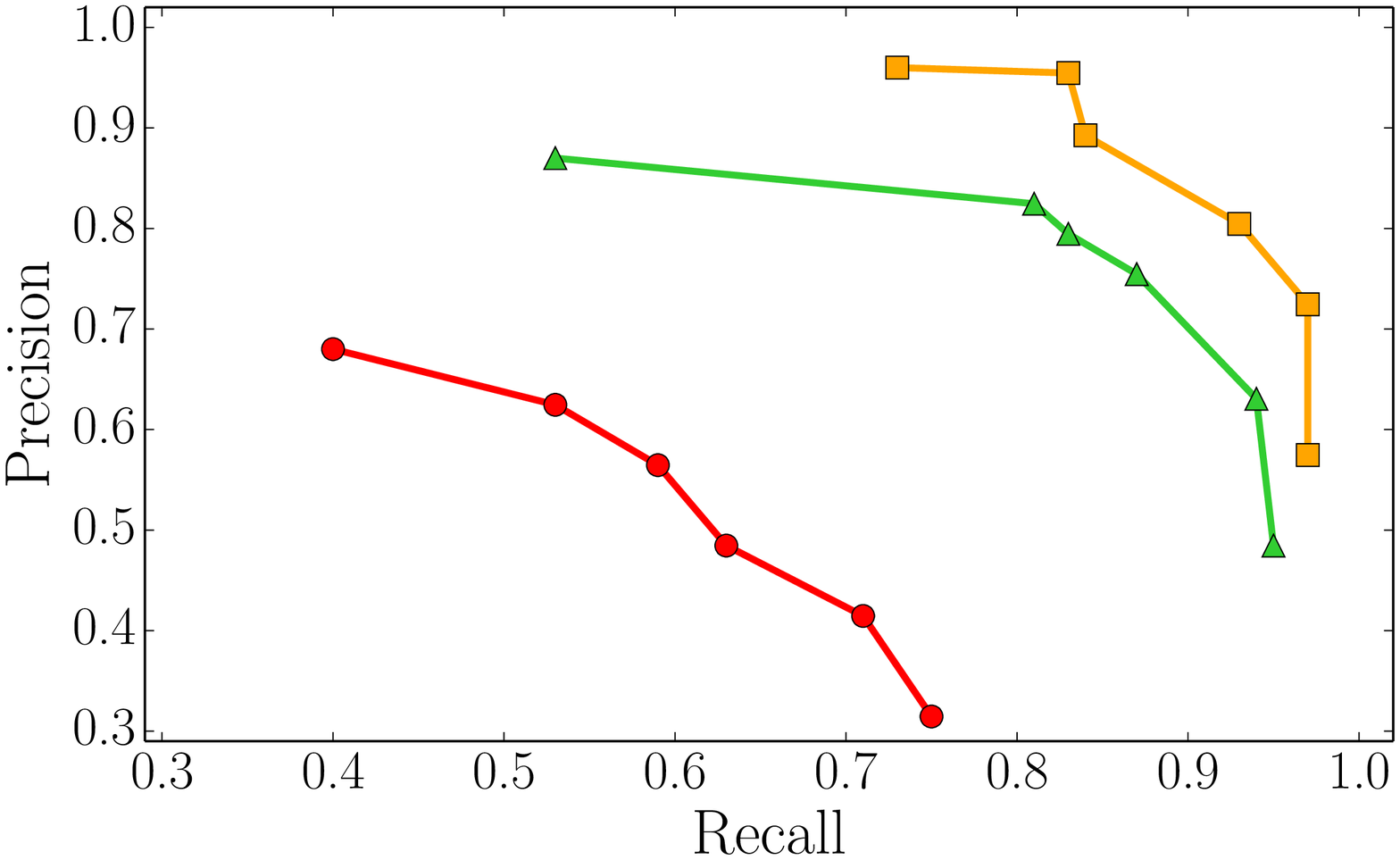}
  \includegraphics[width=0.32\textwidth]
  {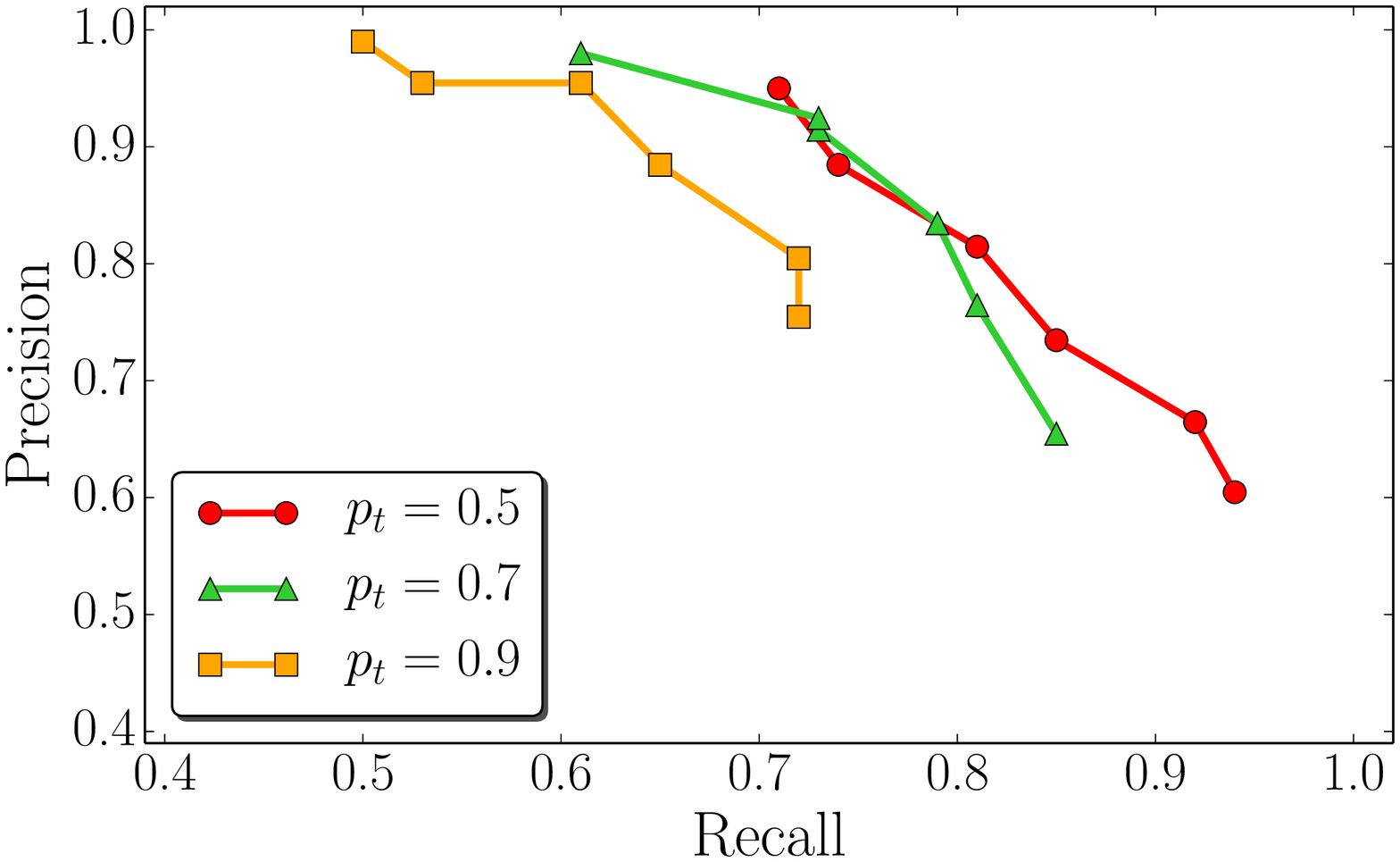}
  \includegraphics[width=0.32\textwidth]
  {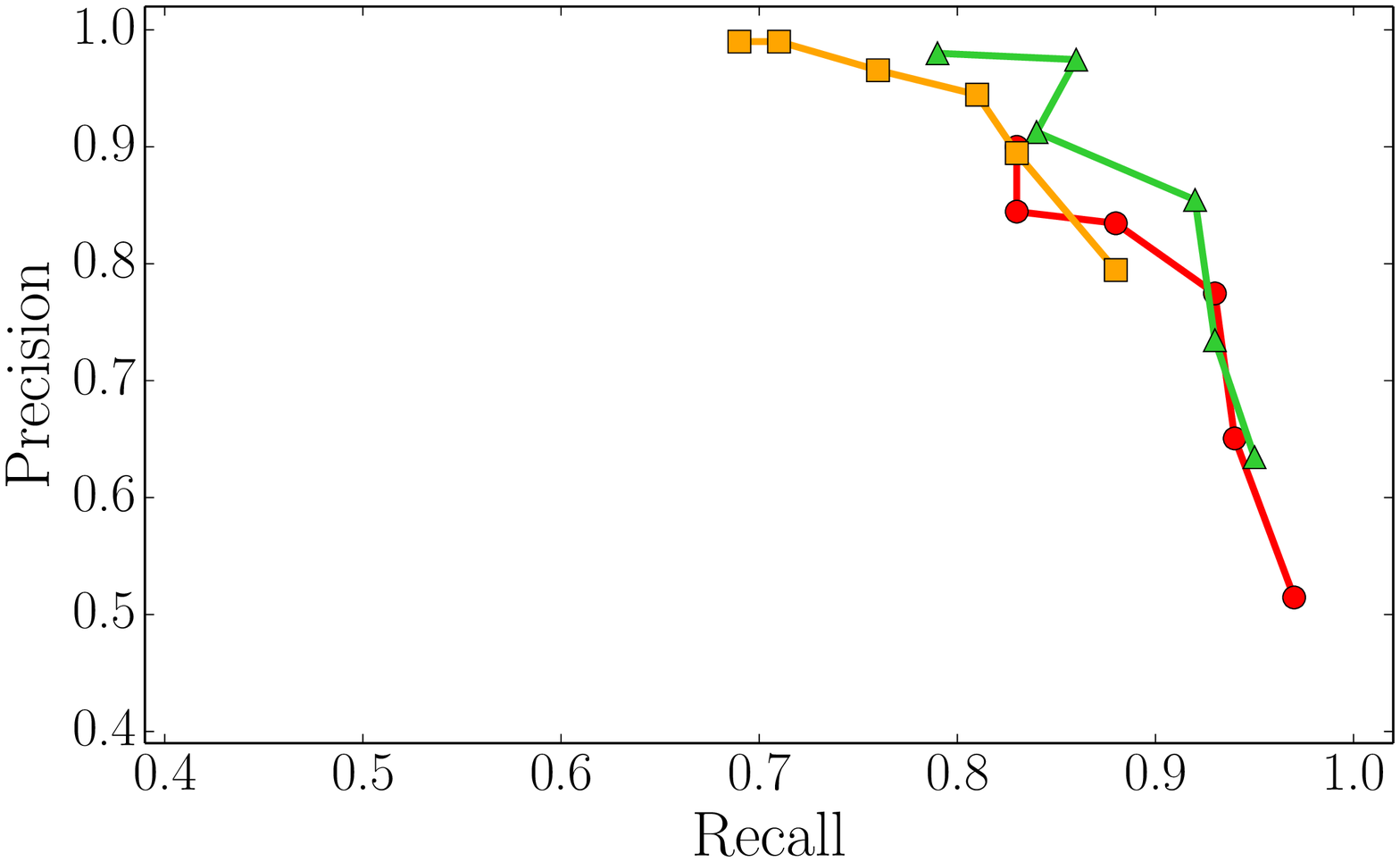}
  \includegraphics[width=0.32\textwidth]
  {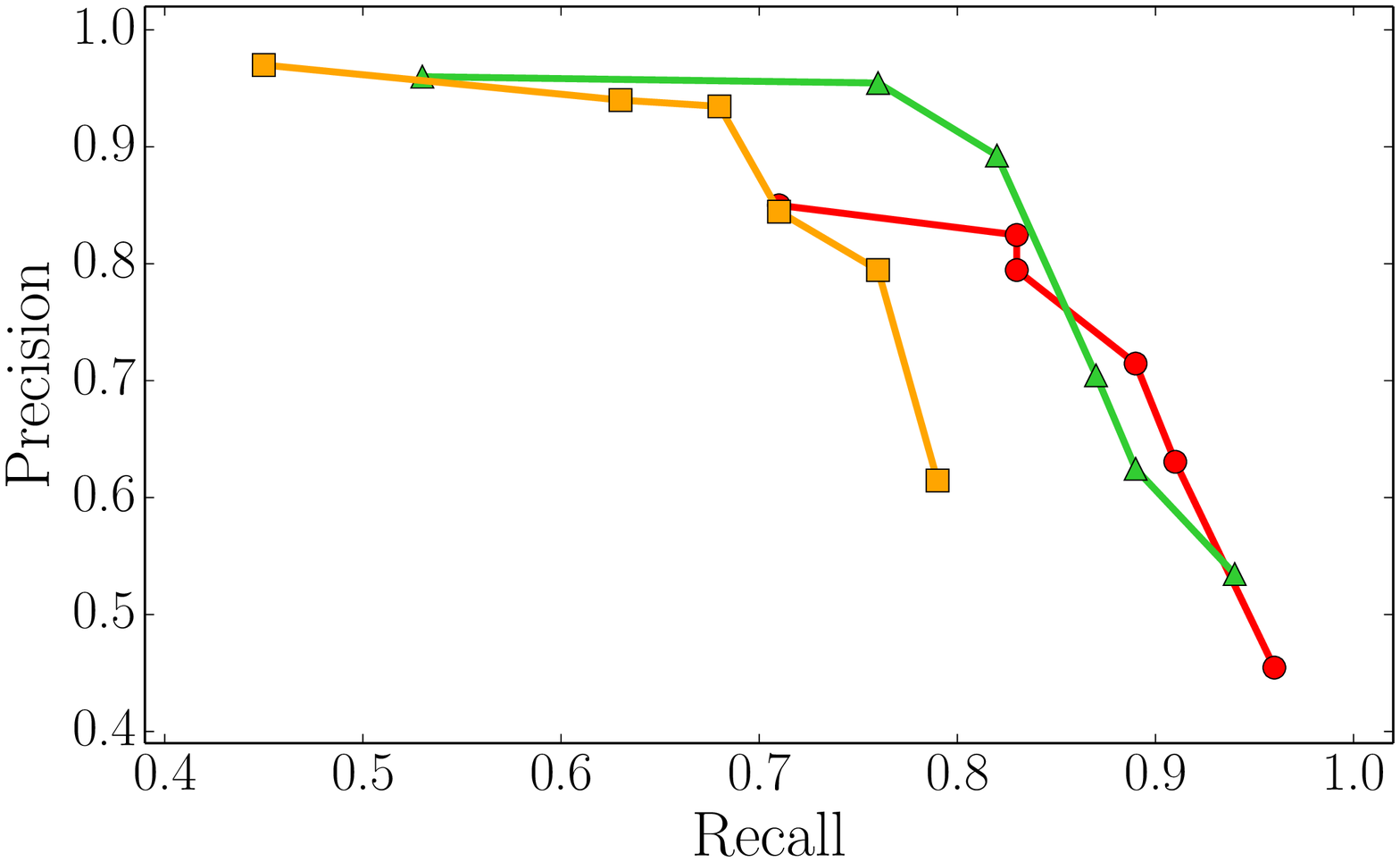}   
  
  \caption{Linkage quality (precision and recall) results of our
    RA-Sig approach with different attribute selection methods and
    varying thresholds $0.5 \le c_t < 1.0$ (top row), different
    numbers of attribute combinations $n_a$ (middle row) in
    Algorithm~1, and different probability thresholds $0.5 \le p_t
    < 1.0$ in Algorithm~2 (bottom row) for different similarity
    thresholds $0.5 \le s_t \le 1.0$.}
  \label{fig:quality}
\end{figure*}

We ran experiments on a server with 64-bit Intel Xeon (2.1 GHz) 
CPUs, 512 GBytes of memory, and running Ubuntu 18.04. We implemented
all techniques in Python (version 3) except \emph{FastLink} for which
we used an Apache Spark based Python implementation developed by
the Ministry of Justice in the United Kingdom~\cite{Splink20}.
Though their original implementation takes the advantage of
parallelisation, we ran all experiments on a single-core
to allow a fair evaluation. We followed~\cite{Abd10,Dem06} to
measure the statistical significance of linkage results. To
facilitate repeatability, the programs and data sets are available
from: \url{https://dmm.anu.edu.au/ra-sig}.

Table~\ref{tab:runtime} shows the runtime of all approaches on the
different data sets. As can be seen, the attribute selection step
of our approach has the longest runtime due to the complexity of the
Apriori based iterative selection method. This runtime can however
be decreased by increasing $c_t$ to reduce the number of candidate
attribute combinations to be processed in each iteration. As shown in
Table~\ref{tab:runtime}, the linkage between two full NCVR snapshots 
took less than 30 minutes to complete all three steps of our approach. 
This shows our approach scales to large database sizes, 
making it applicable for large-scale linkages of databases that 
contain errors and missing values. Overall, \emph{P-Sig} was the 
fastest approach (however it cannot handle errors nor missing values),
while \emph{ProbLink}, \emph{ReDistLink}, and \emph{FastLink} required 
substantially longer runtime due to the weight calculations and
attribute string comparison they employ~\cite{Chr12a}.

The top row of Figure~\ref{fig:quality} shows the linkage quality
of our approach for different values of $c_t$ used in the adaptive 
attribute selection method. A low value of $c_t$ resulted in lower 
linkage quality because attribute combinations that are selected can
include QID attributes with more missing or less distinct values.
However, irrespective of the value of $c_t$, the Apriori based
attribute selection method results in substantially increased
linkage quality compared to randomly selecting QID attribute 
combinations for attribute signatures.


The middle row in Figure~\ref{fig:quality} shows the linkage 
quality of our approach with different numbers of attribute
combinations, $n_a$, in Algorithm~1. As can be seen, an increase 
in the number of combinations, $n_a$, improves both overall
precision and recall. This is because more attribute signatures can
be generated for each record which can potentially identify matching
record pairs even their QID values are missing or contains errors
and variations. 

\begin{table*}[t!]
\centering
\caption{Average precision and recall results (P / R) with
  $s_t=0.8$. The best results are shown in bold.}
\label{tab:quality}
\begin{small}
\begin{tabular}{lcccccc} \hline \noalign{\smallskip}
  ~ & DBLP - Scholar & DBLP - ACM &
  \hspace{3mm}UKCD\hspace{3mm} &
  \hspace{2mm}NCVR-20\hspace{2mm} &
  \hspace{3mm}NCVR-50\hspace{3mm} &
  NCVR-2017 - NCVR-2019 \\
  \noalign{\smallskip}\hline\noalign{\smallskip}
  ProbLink~\cite{Fel69} & 0.53 / 0.31 & 0.94 / 0.53 &
    0.83 / 0.64 & 0.72 / 0.83 & 0.67 / 0.41 & 0.82 / 0.81\\ 
  ReDistLink~\cite{Ong14}~~ & 0.67 / 0.32 & 0.96 / 0.57 &
    0.84 / 0.65 & 0.73 / 0.84 & 0.67 / 0.66 & 0.82 / 0.84 \\
  P-Sig~\cite{Zha18} & \textbf{0.87} / 0.61 &
    0.97 / 0.65 & \textbf{0.95} / 0.71 &
    \textbf{0.97} / 0.82 & 0.91 / 0.70 & 0.95 / 0.85 \\
  FastLink~\cite{Ena19} & 0.73 / 0.58 &
    0.96 / 0.69 & 0.91 / 0.71 &
    0.93 / 0.77 & 0.68 / 0.63 & 0.85 / 0.88\\
      
  \textbf{RA-Sig} &  0.86 / \textbf{0.73}  &
     \textbf{0.99} / \textbf{0.83}  &
     0.92 / \textbf{0.86} &
     \textbf{0.97} / \textbf{0.91} & 
     \textbf{0.96} / \textbf{0.84} & \textbf{0.98} / \textbf{0.93}\\    
    \noalign{\smallskip} \hline 
    
\end{tabular}%
\end{small}
\end{table*}

We measured the linkage quality of our approach with different
probability thresholds $p_t$ for the attribute signature generation
in Algorithm~2. As shown in the bottom row of
Figure~\ref{fig:quality}, a lower $p_t$ value reduces the precision
of our approach because the generated attribute signatures are not
distinct enough (they are shared by many records), which can
potentially lead to false positives. On the other hand, a higher
$p_t$ value tends to increase precision with a drop of recall
because only a few signatures are compared in the matching process.
We also noted that a lower $p_t$ value can increase the runtime of
our approach by 5\% to 10\% due to many attribute signatures are 
being generated in Algorithm~2. 

The average linkage quality achieved by the different approaches for 
similarity threshold $s_t=0.8$ in Algorithm~3 is shown in 
Table~\ref{tab:quality}. As can be seen, \textbf{RA-Sig} resulted
in high precision and recall values with all data sets which
illustrates that the use of relationship information between
records can correctly match records even when their QID attributes
contain missing values, or variations and errors. The linkage
quality results remain qualitatively similar if we change the
threshold $s_t$ from 0.5 to 1.0. 

The probabilistic linkage approaches performed poorly 
due to missing values that resulted in low similarities (even with
weight redistribution~\cite{Ong14} or estimation of weights 
considering missing values~\cite{Ena19}), and many record pairs were
classified as non-matches. \emph{P-Sig} achieved similar
precision as our approach because we are using the same set of 
candidate signatures generated in Algorithm~1 for \emph{P-Sig}. 
However, \textbf{RA-Sig} achieved, based on Tukey's honestly
significant difference (HSD) post hoc test~\cite{Abd10}, statistically 
significant improvement in recall (10\% to 20\%) even if the 
QID attributes contain large numbers of missing values.


\section{Large Scale Real Data Linkage}
\label{sec:german_linkage}

In this section, we describe how we apply our approach in a 
real-world record linkage setting. We use our approach to link two 
simulated Census data sets containing information of all 
households in Germany. 
\smallskip

\noindent\textbf{Problem Description:}
%
%
Due to the number of changes required on population censuses and 
increasing demands for faster and more frequent census statistics,
European countries transition from traditional census operations 
to register-based censuses~\cite{Valente2010}. In Germany, the next
census will be (similar to the Census 2001) a combination of a
register-based and a traditional census. After the Census 2022,
Germany will use a register-based census to decrease respondent
burden, cost and time needed for
processing~\cite{KoernerKrauseRamsauer2019}. A major challenge for 
any register-based census is linking various available administrative
registers if no unique personal identification number is available.

\smallskip
\noindent\textbf{German Census Simulation Data:}
Since no unique personal identification number is currently 
available in Germany, QID attributes such as names, date of birth,
and place of birth have to be used for linking. Data protection 
regulations, such as GDPR ~\cite{GDPR2016}, do not permit the 
permanent storage of QID attribute values in official statistics. 
Therefore, data sets used for the development of linkage applications
have to use artificially simulated
data~\cite{SchnellBorgsWedekind2021}. Such data have
to consider regional variations in household size, compositions of
households, number and kind of institutional populations, name
variations according to date of birth and many more restrictions
usually not considered when generating artificial data for
simulations. Furthermore, the amounts of errors and the frequency
of incomplete QID attributes have to be considered.

Destatis~\cite{Dest}, the Federal Statistical Office of Germany, 
therefore, commissioned a research group at the University of
Duisburg-Essen to develop suitable data sets which fulfil the given
constraints. These data sets are
based on real-world samples of first names and last names, the
actual distribution of households at given addresses and marginal
distributions of population characteristics~\cite{SchnellBorgsWedekind2021}. 
Two of these data sets,
with 81,194,354 and 55,321,209 with an overlap of 55,126,620 records,
are used in the experiments presented here.
We use first and last names, date of birth, gender, and street address as
QIDs for linkage and household identifiers to get relationships
between entities. The two data sets contain 1\% missing values, with
4\%, 3.3\%, 2.6\%, and 1.3\% of individuals having changes in their
first name, last name, date of birth, and street address, respectively.

\smallskip
\noindent\textbf{Results and Discussion: }
Table~\ref{tab:census_linkage} shows the runtime and linkage quality 
results our approach achieved linking these two census data sets. 
As can be seen, our approach completed the linkage in less than 5 
hours. Our approach generated approximately 400 million unique
attribute signatures in Algorithm~2 when we selected the five best 
attribute combinations in Algorithm~1, and set the score weight as
$\alpha=0.5$ and attribute selection threshold as $c_t=0.6$, 
respectively. As shown in Table~\ref{tab:census_linkage}, our 
approach achieved an overall precision and recall of 0.999 and 0.962, 
respectively, which involves the comparison of approximately 54 
million candidate record pairs. 

The initially used combination of algorithms (including exact 
matching~\cite{Chr12a}, phonetic matching~\cite{Chr12a}, and 
FastLink~\cite{Ena19}) needed more than 300 hours on average to 
link the two data sets achieving similar linkage quality which shows
the superiority of our approach when linking large databases 
containing millions of records.

\begin{table}[t!]
\caption{Average runtime (in seconds) and linkage quality results 
 with $s_t=0.8$ for German census data sets.}
\label{tab:census_linkage}
\begin{small}
\begin{tabular}{lll} \hline \noalign{\smallskip}
\multirow{3}{*}{Runtime}  & 
\begin{tabular}[c]{@{}l@{}}Attribute selection step (Algorithm~1)\end{tabular} & 7,903 \\
~ & \begin{tabular}[c]{@{}l@{}}Signature generation step (Algorithm~2)\end{tabular} & 4,573 \\
~ & \begin{tabular}[c]{@{}l@{}}Record matching step (Algorithm~3)\end{tabular} & 3,438 \\ 
 \noalign{\smallskip}\hline\noalign{\smallskip}
\multirow{2}{*}{Linkage quality} & Precision & 0.999 \\
 ~ & Recall & 0.962 \\ 
\noalign{\smallskip}\hline
\end{tabular}
\end{small}
\end{table}


%

\section{Conclusion}
\label{sec:conclusion}

We have presented an efficient approach to link records even
when they contain errors or missing values. Our approach generates
suitable attribute signatures using an unsupervised selection
method, and relational signatures from relationship information
between records. Combined these signatures can uniquely identify
individual records and facilitate very fast and accurate linking of
large databases compared to other state-of-the-art linkage
techniques. As future work, we plan to evaluate our approach under
different missing data scenarios, and to use graph embeddings to
generate improved relational signatures. We also aim to extend our
approach to facilitate efficient and accurate private record
linkage~\cite{Christen2020lsd}.


\balance
\bibliographystyle{ACM-Reference-Format}
\bibliography{arXiv}

\end{document}